\documentclass[
    ,final            
  ]
  {aipproc}

\layoutstyle{6x9}

\begin{document}

\title{Identifying A-stars in the CoRoT-fields IRa01, LRa01 and LRa02}


\author{Daniel Sebastian}{
  address={Th\"uringer Landessternwarte Tautenburg, 07778 Tautenburg, Germany}
}
\author{Eike W. Guenther}{
  address={Th\"uringer Landessternwarte Tautenburg, 07778 Tautenburg, Germany}}

\begin{abstract}
Up to now, planet search programs have concentrated on main sequence
stars later than spectral type F5.  However, identifying planets of
early type stars would be interesting.  For example, the mass loss of
planets orbiting early and late type stars is different because of the
differences of the EUV and X-ray radiation of the host stars.
As an initial step, we carried out a program to identify suitable A-stars
in the CoRoT fields using spectra taken with the AAOmega spectrograph.
In total we identified 562 A-stars in IRa01, LRa01, and LRa02.
\end{abstract}

\maketitle


\section {Introduction}

CoRoT is a satellite mission launched in December 2006. It is
specialized on the detection of extrasolar planets and for studying
the pulsations of stars by means of ultra-precise photometric
measurements. The photometric accuracy achieved is 10 to 100 times
better than what be achieved from the ground ($10^{-4}$ in the
exoplanet channel).  In the so-called long runs, the fields are
observed continuously for 150 days.  About 500 stars are observed with
the full sampling rate of 32 seconds in the exoplanet-channel.  All
other stars, typically about 6000, are observed with a sampling rate
of 8.5 minutes.  Up to now, CoRoT has discovered more than 15
extrasolar planets. Amongst them is the first transiting rocky planet
found outside the solar system (CoRoT-7b), a planet of a young, active
star (CoRoT-2b), a temperate planet (CoRoT-9b), and the first
transiting brown dwarf orbiting a normal star (CoRoT-3b). The CoRoT
objects open up a new window for studying extrasolar
planets. However, the host stars of all these planets are F,G, and
K-stars.

As outlined in detail in Guenther et al. (2010) it would be very
interesting to detect planets of earlier type stars.  For example, the
mass loss of planets orbiting early and late type stars is different
because of the differences in the EUV and X-ray radiation of the host
stars.  By comparing the properties of planets orbiting A-stars and
late-type stars, we can find out what the influence of the central
star for the planet is. However, detecting planets of A-type stars by
means of transit observations is difficult, because the transits are
shallower than for smaller stars. Additionally, many A-stars
oscillate, which makes the analysis of the light curves rather
difficult. In order to make progress in this field of research it is
thus essential to identify the A-star first. This is the aim of this
work.

\section {Identifying candidates and observations}

Prior to the launch of the satellite, many CoRoT fields were observed
using multi-colour photometry (B,V,R,I). Almost all stars that CoRoT
observed are also in the 2MASS data-base, so that J,H,K magnitudes are
available.  The whole photometric data is accessible threw EXODAT
(Deleuil et al. 2006).

However, identifying A-type stars only on the basis of the broad-band
photometry is difficult, because of the reddening. We thus take a two
step approach. In the first step we pre-select the stars based on
photometry. As a criterion we use the B-V colors, and select all stars
with $B-V= -0.16^m$ to $0.42^m$, corresponding to an un-reddened B5V
to F5V star (Binney \& Merrifield, 1998). The second step is then the
proper determination of the spectral types based on spectroscopy.
While the colour-selection approach reduces dramatically the number of
stars that we have to study, we may loose a few highly reddened
A-stars in this process. Thus, our survey does not aim in detecting
{\em all} A-stars observed by CoRoT but it aims in finding a sample of
A-stars that can be studied.

Although we preselected the targets, we still have to take spectra of
several hundred stars. Luckily, as part of the ground-based follow-up
observations the CoRoT fields IRa01, LRa01, and LRa02 were observed
with the multi-object spectrograph AAOmega mounted on the AAT
(Anglo-Australian-Telescope when the observations were taken, now
renamed to Australian-Astronomical-Telescope).  AAOmega is ideal for
our purposes, as this instrument allows to take spectra with up to 350
stars in a field of 2\textdegree $\times$ 2\textdegree (Saunders et
al. 2004; Smith et al. 2004). The CoRoT fields IRa01 and LRa01 have a
size of 1.4\textdegree $\times$ 2.8\textdegree, and LRa02
1.4\textdegree $\times$ 1.4\textdegree.  Mounted in the prime focus of
this telescope is a fibre positioner that feeds the AAOmega
spectrograph. Using the AAT together with the AAOmega spectrograph we
have obtained more than 20\,000 spectra of stars in the CoRoT-fields.

\begin{figure}[h!]
  \includegraphics[height=.25\textheight]{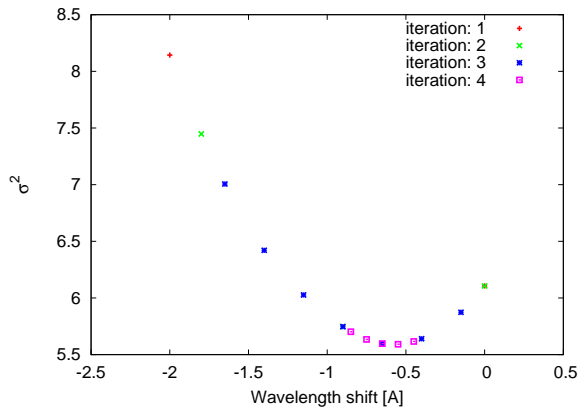}

\caption{The spectral types are obtained iteratively by fitting the
  observed spectra to templates by minimizing $\sigma^2$. Shown here
  is the derived $\sigma^2$ vs. the shift in wavelength. The best
  match is obtained at the minimum of $\sigma^2$.}
  \label{Abb:1}
\end{figure}

\begin{figure}[h!]
  \includegraphics[height=.3\textheight]{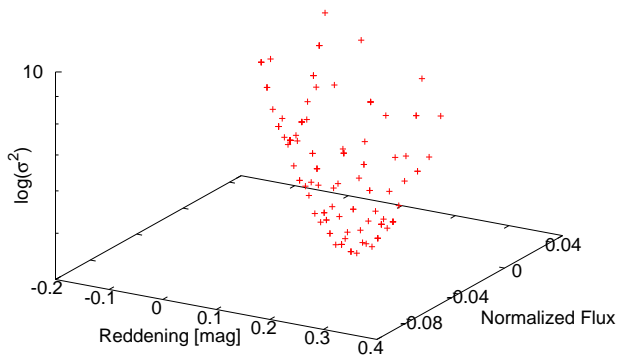}
\caption{Same as Fig. 1 but for the flux and the reddening.}
  \label{Abb:2}
\end{figure} 

IRa01, LRa01, and LRa02 are located in the so-called anti-center
``eye`` of the CoRoT-mission (RA 6h to 7h \& DEC -10\textdegree\,to
10\textdegree).  The data was obtained in two campaigns. The first
campaign was carried out from the $13^{th}$ to the $20^{th}$ of
January 2008.  Unfortunately, observations could only be obtained on
the $13^{th}$ and $14^{th}$ of January.  Nevertheless, 4112 spectra of stars in
IRa01 and LRa01 were taken. The second campaign was carried out from
the $28^{th}$ of December 2008 to the $4^{th}$ of January
2009. Observations were carried out in all eight nights, and we took
spectra of 14187 stars in all three fields.

We used ``Configure'', the target allocation software in order to find
the optimum configuration of the fibres. In each setting we typically
managed to place 350 fibres onto target stars, and 25 fibres onto the
sky background. In order to optimize the exposure time were minimized
the spread in brightness of the stars observed in each setting. We
started our observations with the fields containing the brightest
stars and subsequently used setting of fainter stars.

Our targets have V-magnitude in the range 10 to 15, corresponding
to the bright part of the CoRoT/Exoplanet targets. Down to mv = 14.5, our
observation cover essentially all stars in the CoRoT-fields.

As usual, a fibre bundle placed onto a relatively bright star for
guiding purposes. In order to monitor any possible field rotation,
typically 6 fibres were place onto stars within the FOB. For the
observations we used the AAOmega spectrograph with the 580V grating in
the blue arm and the 385R in the red arm.  The spectra cover the
spectral range from 3740 to 5810 \AA \, in the blue arm, and 5650 to
8770 \AA \, in the red arm. The resolution is R=1300.  

Each field was observed for 30 to 45 minutes. In order to avoid
any saturation, and in order to make the removal of cosmic rays easy,
we split the observing time spend on each field into three or more
exposures.

All calibration frames (flat, arcs and bias-frames) were taken as it
is common practice with AAOmega: Bias frames in the afternoon before
each observing night, flats and arcs before the observations of each
field. The sky subtraction is not critical, because we observed only
stars brighter than 15.0 mag and the observations were carried out
during dark time.  Nevertheless, for subtracting the spectrum of the
night sky from the stellar spectra, we have to calibrate the relative
throughput for each fibre. Because the throughput varies depending on
the bending of the fibre, these measurements had to be done after the
fibres have been positioned.  The throughput of each individual fibre
was determined by taking spectra during dawn, or observing fields of
blank sky during the night.  The spectra were reduced using the
2dfrdr-reduction package which has especially been developed for
AAOmega.

\section{How the spectral types are determined}

As outlined above, we take a two-step approach. In the first step. we
select suitable candidates based on their B-V-colours. The second step
then is the determination of the spectral type using the AAOmega
observations.

This is done by deriving which spectrum from a library of template
spectra matches best the observed spectrum. We use ``The
Indo-U.S. Library of Coud\'{e} Feed Stellar Spectra''(Valdes et
al. 2004) for this purpose. For each template $\sigma^2$ is calculated
as the sum of the squared differences between the template and the
observed spectrum. In order to match each template to the observed
spectrum, we first shift the spectrum to the correct position in
wavelength, adjust the flux, and remove the extinction. This process
is done iteratively, always by varying each of these parameters and
then minimizing $\sigma^2$. In this way we automatically determine the
radial velocity and the extinction, $A_V$ (Binney \& Merrifield 1998;
Fig.1, 2) for each star.  Since the library includes templates of
different luminosity classes, we can also determine the luminosity
class for the star. (Fig.3).

\begin{figure}
  \includegraphics[height=.3\textheight]{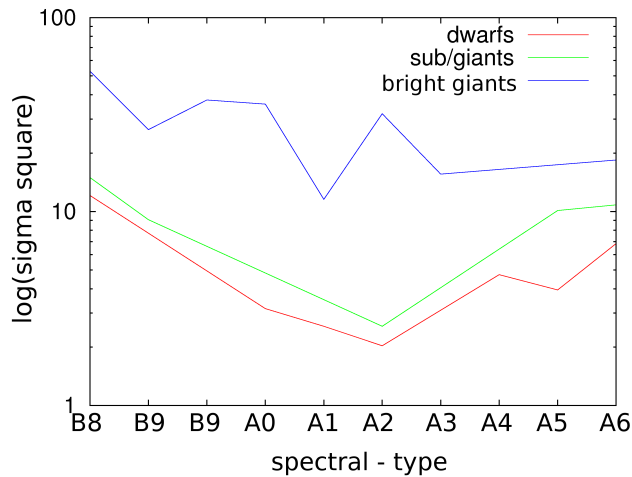}
  \caption{Shown is the differences between the templates and the observed
    spectrum expressed in $\sigma^2$ for templates of different
    spectral types and luminosity classes.  The minimum of $\sigma^2$ is
    achieved for A2V.  }
  \label{Abb:3}
\end{figure}

\section{The accuracy of the method}

Fig. 4 shows an observed spectrum together with the best matching
template. The red line is the observed spectrum of the star after
correcting it for extinction, radial velocity, and removing also the
off-set in flux. The green line is the template spectrum of an A5V
star.  The templates matches the observed spectrum extremely well.

\begin{figure}[h!]
  \includegraphics[height=.3\textheight]{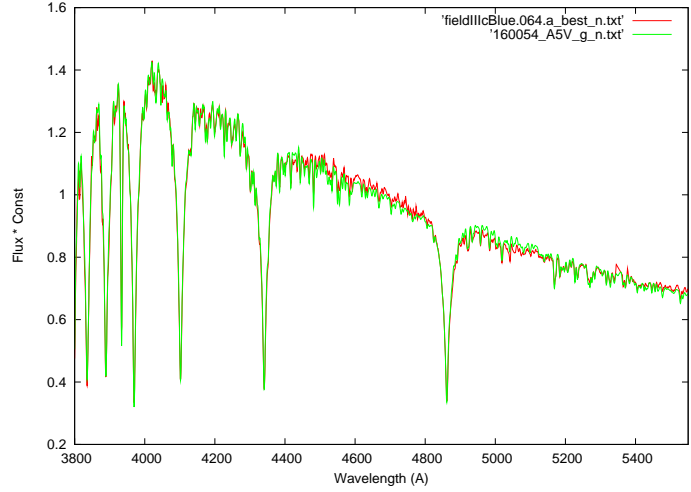}
  \caption{Observed spectrum (red line) and the best matching template (green line) for an A5-star}
  \label{Abb:4}
\end{figure}

The question however is, how accurate the method for determining the
spectral types is. For 178 stars, we have obtained several spectra.
This data thus allows a thorough test how accurate our method is.  In
these cases, we derive the spectral type of the same star several
times using different spectra. The differences of the spectral types
derived for the same star thus gives us the error of the
measurements. The result is shown in Figure 5. A value of 1.0 means that
the difference of the spectral type of two determinations of the same
star is one subclass. We find that our error is $1.13\pm0.08$
subclasses. As can be seen in Figure 5, there are a few outliers. These
are largely due to the fact that a few, individual spectra were
effected by instrumental problems.  As shown in Figure 3, we can also
reliable distinguish between dwarfs and giants. Since the difference
between dwarfs and sub-giants is rather small, distinguishing these is
less certain.

It turns out that the main limitation of the determination of the
spectral types is not the quality of the spectra, or the analysis of
the data but the quality of the templates taken from the
literature. The libraries published by different authors gave slightly
different results (Le Borgne, J.-F. et al. 2003 , and Jacoby, Hunter
and Christian).

\begin{figure}
  \includegraphics[height=.25\textheight]{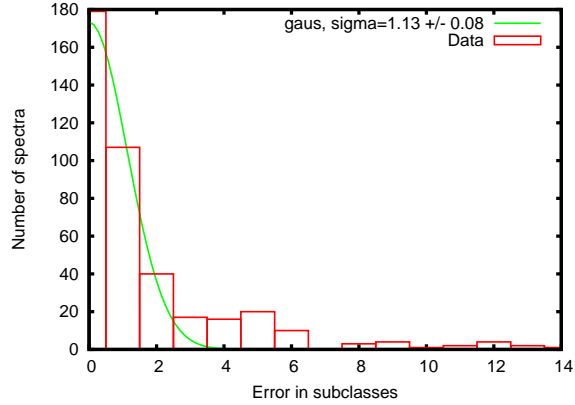}
  \caption{The error in subclasses, derived from the difference
           of the spectral types obtained for stars of which
           several spectra were taken.}
  \label{Abb:5}
\end{figure}

\begin{figure}
  \includegraphics[height=.25\textheight]{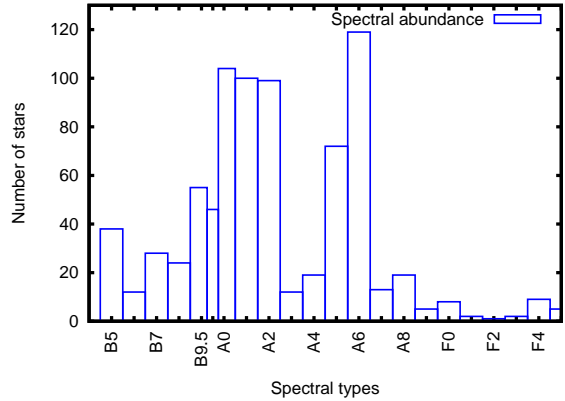}
  \caption{Distribution of spectral types in our sample.}
  \label{Abb:6}
\end{figure}

\section{The spectral types derived}

As already mentioned above, we selected stars according to their B-V
colours. As a selection criteria, we used $B-V= -0.16^m$ to $0.42^m$,
corresponding to an un-reddened B5V to F5V stars (Binney \&
Merrifield, 1998). In total 805 stars were observed with AAOmega which
matched this criterion. After the detailed analysis of the spectra we
identified 562 A-stars in IRa01, LRa01, and LRa02. Thus, $\sim70\%$ of
the stars in the range between $B-V= -0.16^m$ and $0.42^m$ are
A-stars.  Figure 6 shows the distribution of spectral types. An
interesting feature is that stars with spectral types of A3V and A4V
are missing.  This does not mean that stars of a certain temperature
do not exist but it means that almost no stars match the A3V and A4V
templates from the literature. This is caused by the slightly odd
definition of the sub-classes in this spectral regime.  The absence of
these types of stars is already well know.

After identifying 562 A-stars in IRa01, LRa01, and LRa02, we now
intend to analyze the CoRoT light-curves of these stars in detail in
order to detect the shallow transit of planets orbiting these stars.

\bibliographystyle{aipproc}   

\begin{theacknowledgments}
We are grateful to the user support group of AAT for all their help
and assistance for preparing and carrying out the observations.  We
would like to particularly thank Rob Sharp, Fred Watson and Quentin
Parker. The authors thank DLR and the German BMBF for the support
under grants 50 OW 0204, and 50 OW 0603.
\end{theacknowledgments}

\end{document}